\documentclass[final,1p,times]{elsarticle} 
\usepackage{graphicx}
\usepackage{amssymb} 
\usepackage{amsthm} 
\usepackage{lineno}
\usepackage{subfigure} 	
\newcommand{\pT}{$p_{\rm T}$}

\newcommand{\ptrig}{$p_{\rm T}^{\rm trig}$}
\newcommand{\pone}{$p_{\rm T}^{\rm trig,1}$}
\newcommand{\ptwo}{$p_{\rm T}^{\rm trig,2}$}
\newcommand{\pthree}{$p_{\rm T}^{\rm trig,3}$}
\newcommand{\pfour}{$p_{\rm T}^{\rm trig,4}$}
\newcommand{\ptjet}{$p_{\rm T,jet}^{\rm ch}$}

\newcommand{\s}{$\sqrt{s}$}
\newcommand{\sNN}{$\sqrt{s_{\rm NN}}$}

\newcommand{\pp}{pp}
\newcommand{\PbPb}{Pb--Pb}

\journal{Nuclear Physics A} 

\begin{document}

\begin{frontmatter} 

% Your Title - please insert
\title{Jet structure in 2.76 TeV \PbPb\ collisions at ALICE}
%% Single author (and collaboration) - please insert
\author{Leticia Cunqueiro (for the ALICE\fnref{col1} Collaboration)}
\fntext[col1] {A list of members of the ALIE Collaboration and acknowledgements can be found at the end of this issue.}
\address{Istituto Nazionale di Fisica Nucleare,Via E.Fermi 40 , Frascati}

\begin{abstract} 
  We present the analysis of the semi-inclusive distribution
  of reconstructed charged particle jets recoiling from a high
  \pT\ hadron trigger in central \PbPb\ collisions at
  \sNN\ $=2.76$ TeV. We measure, subtract  and unfold
  the large underlying event background in such collisions, utilizing
  a novel technique that does not impose fragmentation bias on the
  measured jet population. The \PbPb\ measurements are compared 
  to a \pp\ PYTHIA reference distribution generated at the same \s\ .
  Modification of jet structure due to quenching is
  explored by varying the cone radius $R$ ($0.2$, $0.4$) and the lower 
  \pT\ cutoff of charged particle constituents ($0.15$, $2.0$
  GeV/$c$). 
   
\end{abstract} 

\end{frontmatter} % do not change

%% linenumbers are useful for reviewing process
%\linenumbers

%\section{Introduction}

The radiation pattern of jets produced in heavy ion collisions is
expected to be modified with respect to vacuum fragmentation, due to interaction with
the medium. Measurement of fully reconstructed jets aims to capture
the full dynamics of jet quenching, to understand the mechanisms of
in-medium jet energy loss and to infer the transport properties of the
medium iself. In these proceedings, the structure of jets in \PbPb\ collisions at $\sqrt{s_{NN}}=2.76$ TeV is studied and compared to vacuum fragmentation, as given by a PYTHIA calculation at the same energy. The measurement is based on the semi-inclusive distribution of reconstructed charged particle jets (anti-$k_{\mathrm T}$, $R=0.2$ and $0.4$) recoiling from a high \pT\ trigger hadron (``h+jet''). The h+jet coincidence measurement enables full correction for the large underlying event background without imposing bias on jet fragmentation, down to low jet \pT.

%\section{The h+jet coincidence measurement}

The measurements are carried out using \PbPb\ collision data recorded
by ALICE in 2010. Jets are reconstructed using the FastJet anti-$k_{\mathrm T}$
algorithm \cite{Cacciari:2011ma} with resolution parameter $R=0.2$ and
$R=0.4$, and with input consisting of all of charged tracks
reconstructed in the ALICE central barrel acceptance ($|\eta|<0.9$,
\pT$>0.15$ GeV/$c$). Jet reconstruction utilizing tracks with \pT$>2.0$
GeV/$c$ was also considered, to study the contribution of soft particles
to the jet energy. All jets with jet axis lying within $|\eta(\rm jet)|<0.5$ are
accepted for further analysis.  The 20\% most central \PbPb\
collisions are selected. In each event, the
charged hadron with the largest \pT\ is designated the ``trigger''
hadron, with momentum \ptrig\ .  The semi-inclusive distribution of
recoil jets is measured by counting the number of jets in the event
population within the recoil azimuth relative to the trigger
direction, $|\varphi(\rm trig)-\varphi(jet)|<\pi-0.6$, binned differentially in
both \ptrig\ and \ptjet\ and normalized to the
corresponding number of triggers. Hadron triggers are considered only
for \pT$>10$ GeV/$c$. The probability of two such triggers occuring in a
central \PbPb\ event is negligible. Such triggers therefore
isolate a single hard process in the collision, for which we aim to
measure the recoil jet distribution. Hadrons are prefered to jet
triggers in this analysis, since measurement of high \pT\ hadrons in
both \pp\ and \PbPb\ collisions can be carried out accurately without
having to correct the measured \pT\ for underlying event effects,
allowing a more precise comparison of the two systems.

The effects on the recoil jets of the large underlying event in central \PbPb\ collisions
are addressed in two distinct steps, correcting separately for the
median background level and for its fluctuations. The first step, carried out on an event-by-event and jet-by-jet basis, 
estimates the median momentum density $\rho$  \cite{Abelev:2012ej} and subtracts $\rho\times{A}$ from
each jet, where $A$ is the jet area \cite{Cacciari:2007fd}.  The $\rho$
calculation excludes the two hardest clusters in the event, to reduce
the influence of the true signal on the background estimate.

Since $\rho$ is calculated as the median of $p_{\rm T}/A$ for all $k_{\mathrm T}$
clusters in the event, about half of the reconstructed jet population
will in practice have \ptjet\ $<0$ after subtraction of $\rho\times{A}$. A
large fraction of these negative energy ``jets'' is due to the recombination of hadrons that are nearby in
phasespace but have been generated by multiple incoherent processes
(``combinatorial jets''). This component of the jet spectrum
contains valuable information about the fluctuation structure of the
events, which also distorts the measurement of true, hard coincidence
jets. We therefore retain this noise component until a late stage in
the analysis, utilizing it to correct the effect of fluctuations
without imposing bias on the hard jet population.
\vspace{-4cm}
\begin{figure}[h]
\begin{center}
\includegraphics[width=0.48\textwidth]{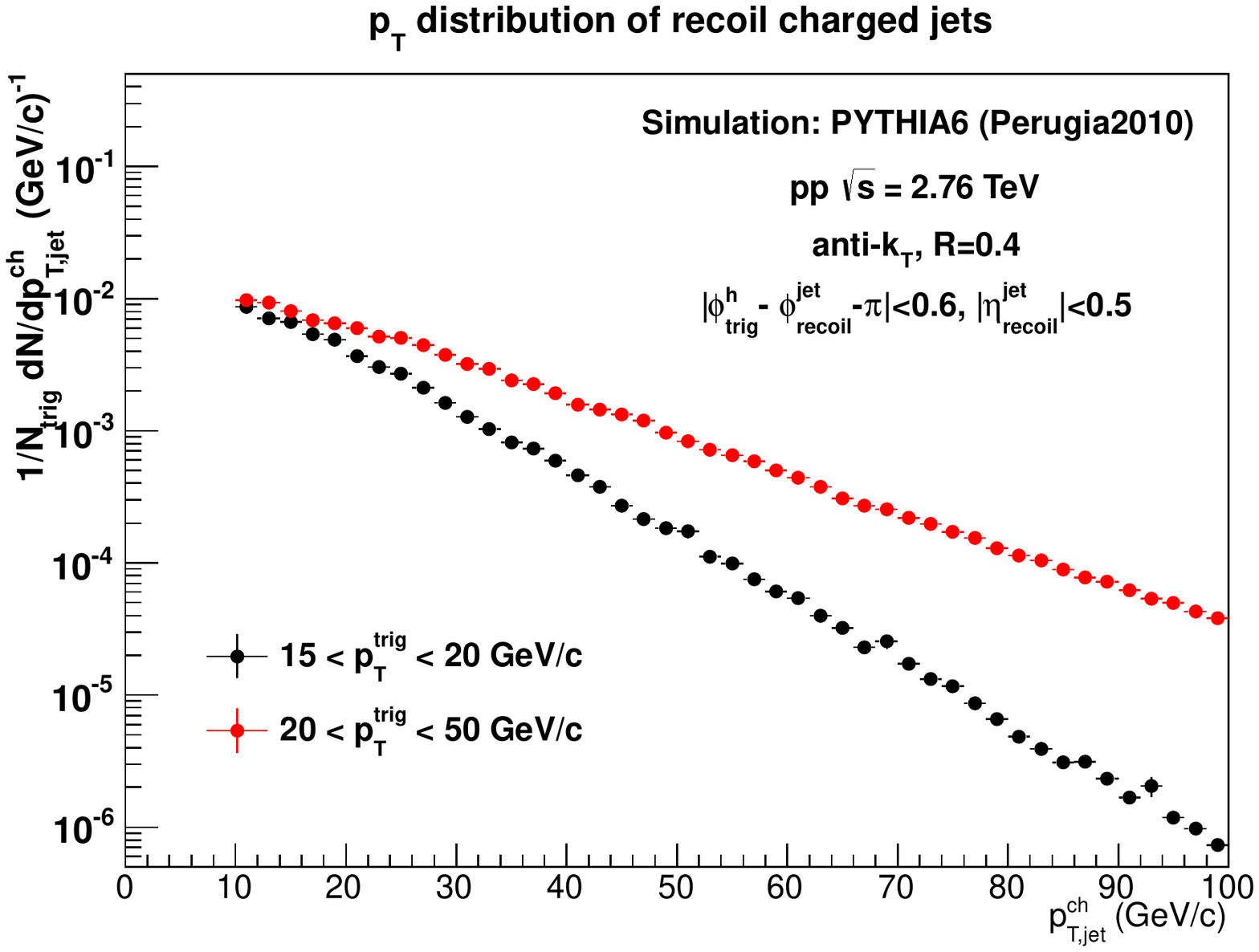}
\includegraphics[width=0.48\textwidth]{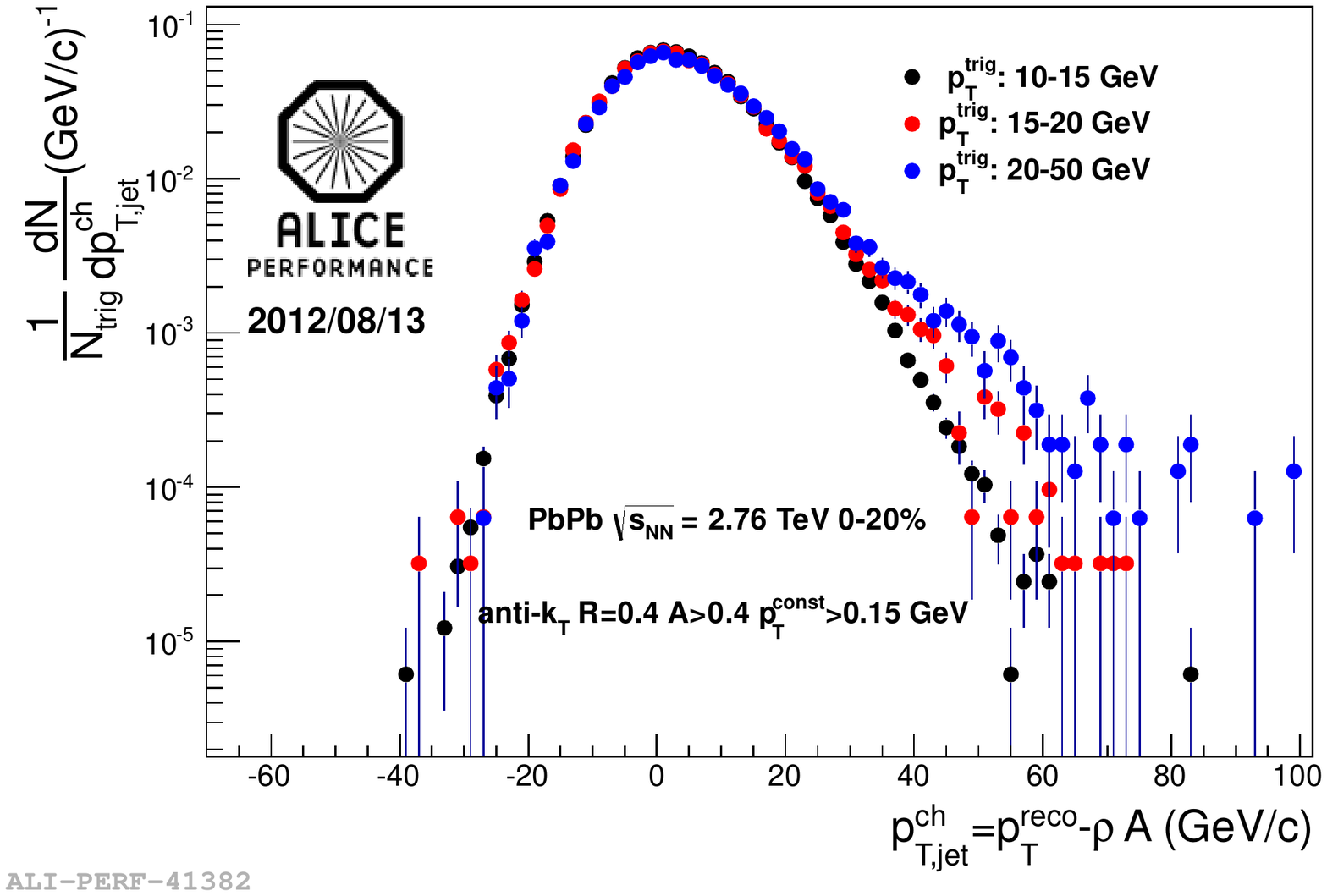}
\end{center}
\caption{Recoil jet distributions for a \pp\ PYTHIA calculation at $2.76$ TeV (left) and the same observable measured in 0-20$\%$ central \PbPb\ collisions }
\label{fig:figref}
\end{figure}
%---

Figure \ref{fig:figref}, left panel, shows a PYTHIA calculation
(Perugia10 tune \cite{Skands:2009zm}) of the semi-inclusive recoil jet spectrum in 2.76 TeV \pp\
collisions for
two different \ptrig intervals. The recoil jet distribution is seen to
depend strongly on \ptrig, as expected, since a harder hadron trigger
biases towards higher $Q^2$ processes on average.

Model studies show that a high \pT\ hadron trigger induces a
geometrical bias, towards jets generated on the surface of the
fireball and directed outward. The jet population recoiling from such
a trigger is therefore biased towards larger in-medium path length
compared to the inclusive population or the population recoiling from
a jet trigger \cite{Renk:2012hz}. The hadron trigger imposes a low
\pT\ suppression on the recoil jet spectrum: at LO the hadron \pT\
provides the strict lower bound for recoil jet \pT, though there are
higher-order and non-perturbative effects that smear this threshold in
practice. There are additional biases induced by the hadron trigger,
in event centrality, reaction plane orientation, and nuclear $k_{\mathrm T}$
effects. However, for the high $p_T$ range  of hadron triggers considered in this
analysis, all such effects have only weak, if any, dependence on
hadron \pT, and their effects are to a large extent removed by the
differential spectrum technique described below.  

In this analysis we apply a new approach to suppress the
combinatorial background for the h+jet measurement
\cite{deBarros:2012ws}. We observe that the distribution of
combinatorial background jets is, by definition, {\it uncorrelated}
with trigger \pT, and its shape is observed to be identical for all
choices of trigger \pT. This raises the possibility of a purely
data-driven elimination of the combinatorial jet population, by
considering the measurement of the {\it difference} of two measured
jet distributions with hadron triggers in different \pT\
intervals:\\ $\Delta_{\rm recoil}(($\pone-\ptwo$)-($\pthree-\pfour$)) =\frac{1}{\rm{N}_{\rm trig}}\frac{\rm{dN}}{\rm{d}p_{\rm T,jet}^{\rm ch}}($\ptwo$<$\ptrig$<$\pone$)-c \frac{1}{\rm{N}_{\rm trig}}\frac{\rm dN}{\rm{d}p_{\rm T,jet}^{\rm ch}}($\pfour$<$\ptrig$<$\pthree$)$

Scaling of the lower $p_{T}^{trig}$ distribution is applied to account
for the observed strict conservation of jet density in the
experimental acceptance, which results in increasing displacement of
combinatorial jets by true, hard coincidence jets as $p_{T}^{trig}$
increases. The scaling factor, $c$,  is measured in the region \ptjet\ $<0$ where the combinatorial
contribution dominates, and differs from unity by about 5\%.

The remaining distribution in this difference observable, is therefore not due to combinatoric jets. It
represents the {\it evolution} of the true recoil jet distribution
from the same hard interaction as the trigger, as the trigger \pT\
evolves from the lower \pT\ trigger interval (``reference'') to the
higher \pT\ trigger interval (``signal''). This observable, while uncommon, is nevertheless perturbatively
well-defined. 

Monte Carlo calculations of this observable require, in addition to
perturbative processes, an accurate description of inclusive particle
production to model the trigger. Comparison to these data then tests
the modeling of the recoil (quenched) jet. We note that the analysis
procedure removes the combinatorial jet component from the measured
jet distribution without imposing any bias on jet fragmentation.

Correction of the measured jet \pT\ for fluctuations utilizes a response matrix based on the
random cone technique \cite{Abelev:2012ej}, while the detector
response is simulated using PYTHIA-generated jets. We unfold using two different
iterative methods, based on $\chi^{2}$ minimization and on Bayes's Theorem. Their difference contributes to the anti-correlated shape uncertainty in the figures.
\vspace{-4cm}
\begin{figure}[h]
\begin{center}
\includegraphics[width=0.48\textwidth]{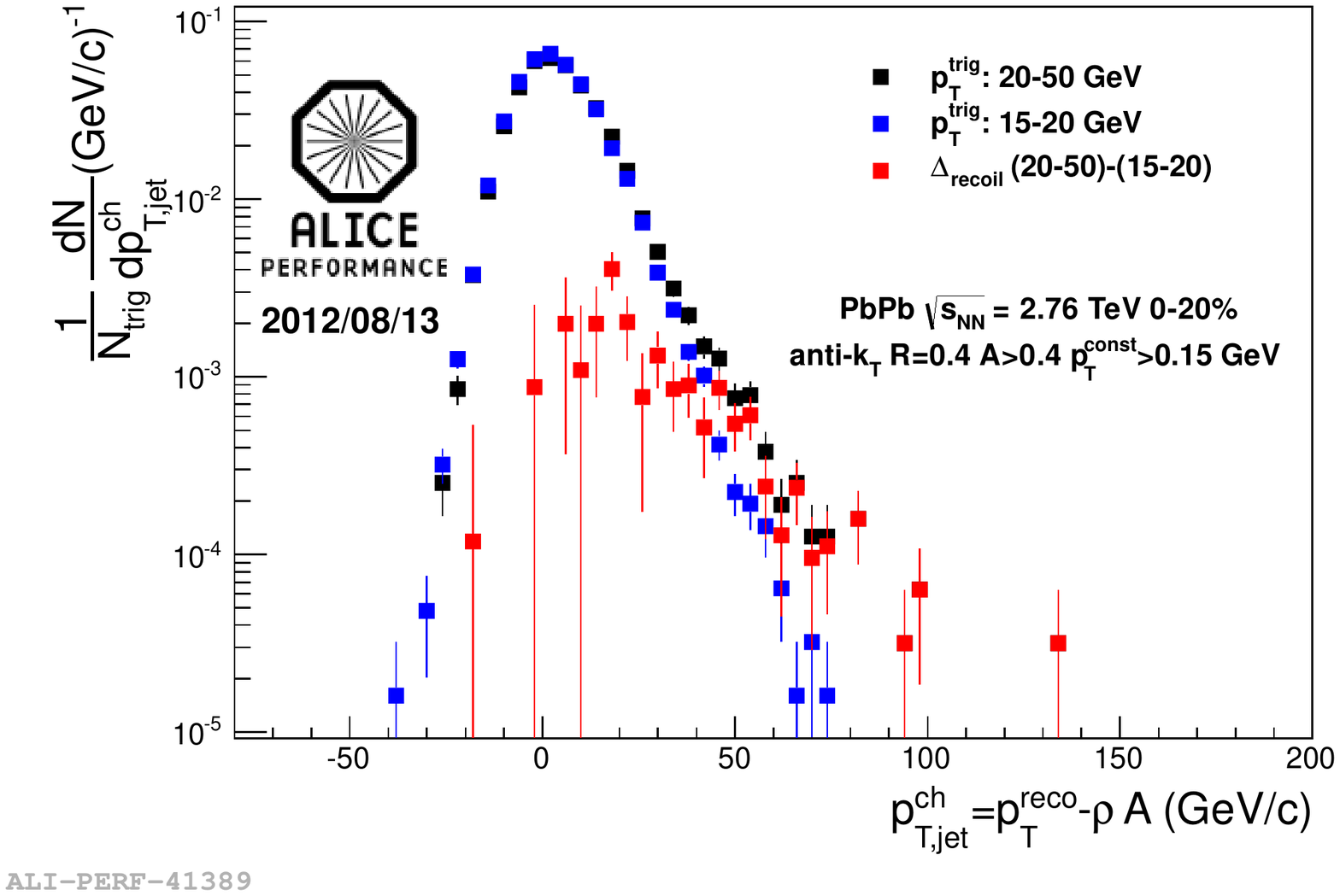}
\includegraphics[width=0.48\textwidth]{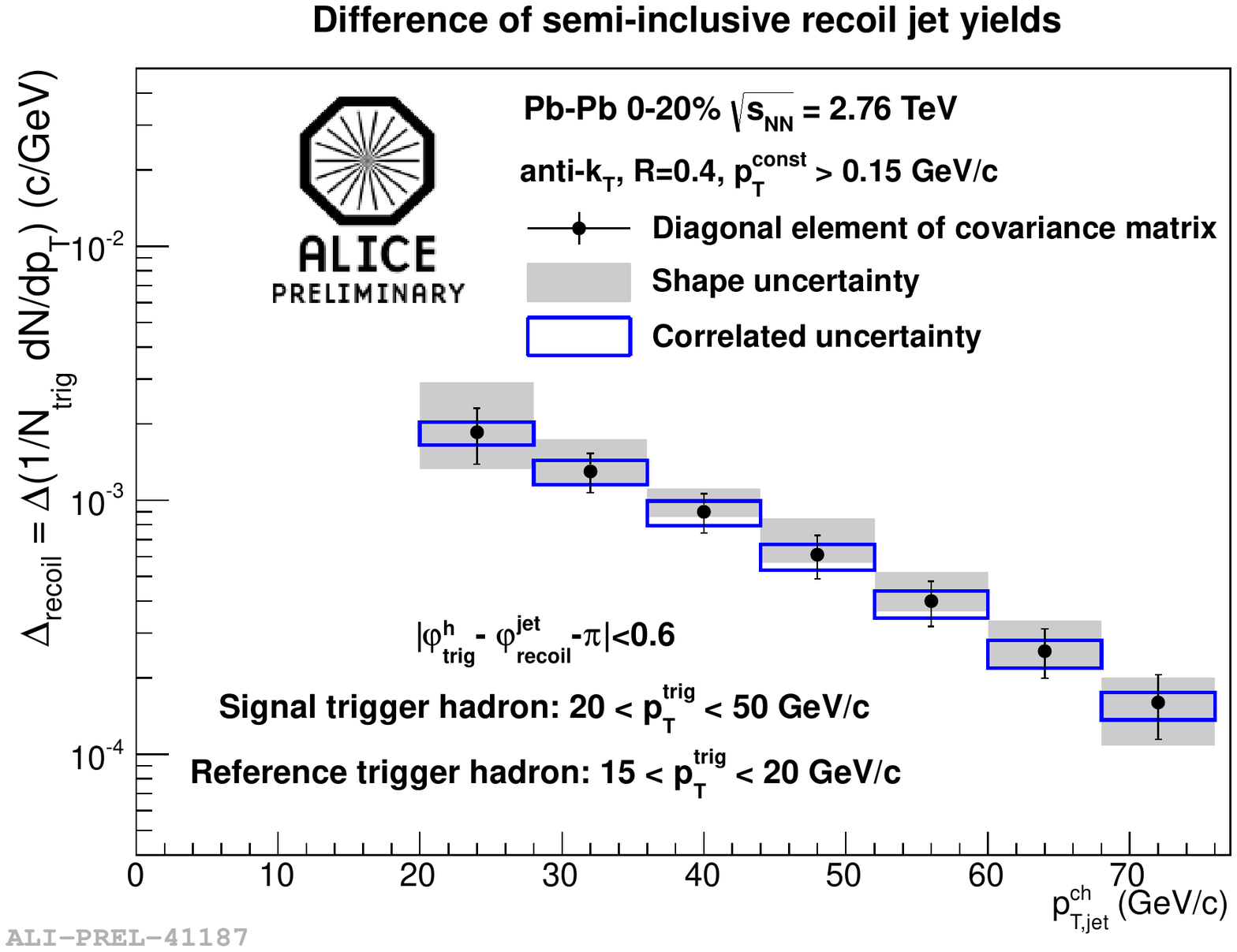}
\end{center}
\caption{$\Delta_{recoil}$((20-50)-(15-20)) raw (left) and corrected (right) distribution for $R=0.4$ and $p_{T}^{const}$=0.15 GeV}
\label{fig:spectrum}
\end{figure}
%---

Fig.\ref{fig:figref} right, shows the h+jet recoil distribution for
$R=0.4$, and for three exclusive intervals of trigger \pT. For
\ptjet$<20$ GeV/$c$ the distributions are uncorrelated with the trigger
\pT\, and are seen to be very similar, consistent with dominance by the combinatorial
jet contribution. In contrast, at larger \ptjet\ the distributions are
correlated with trigger \pT\ , consistent with dominance by hard jets
from the same hard scattering as the trigger. \\
In Fig.\ref{fig:spectrum} (left) the recoil jet distribution for two trigger classes as well as their difference $\Delta_{\rm recoil}(($20-50$)-($15-20$))$ are plotted. Note the exponential shape of the distributions, indicating that the unfolding of background fluctuations and detector effects will be a small correction.
The plot on the right shows the corrected $\Delta_{recoil}$ for the same choice of trigger classes.

\begin{figure}[h]
\begin{center}
\includegraphics[width=0.48\textwidth]{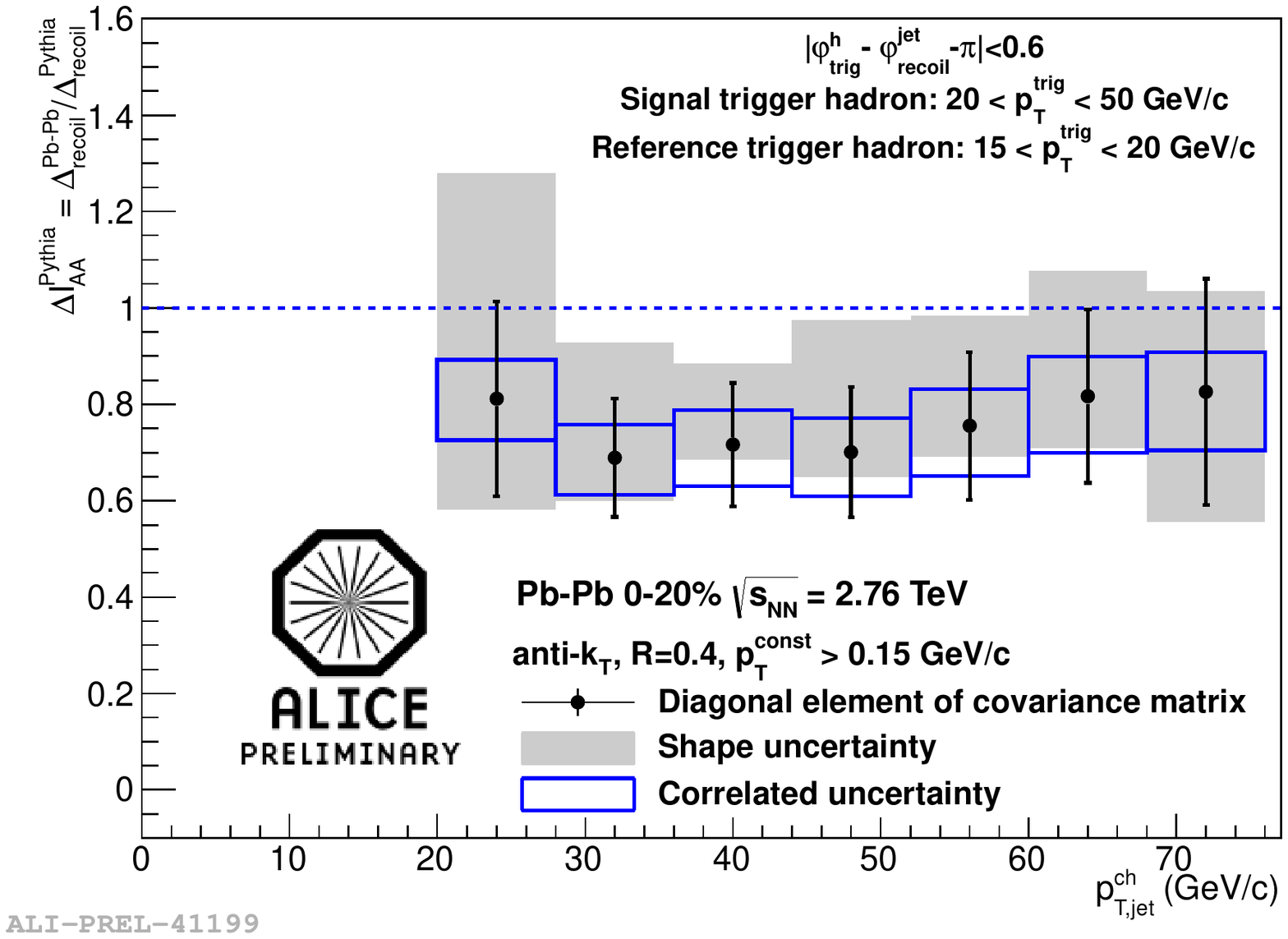}
\includegraphics[width=0.48\textwidth]{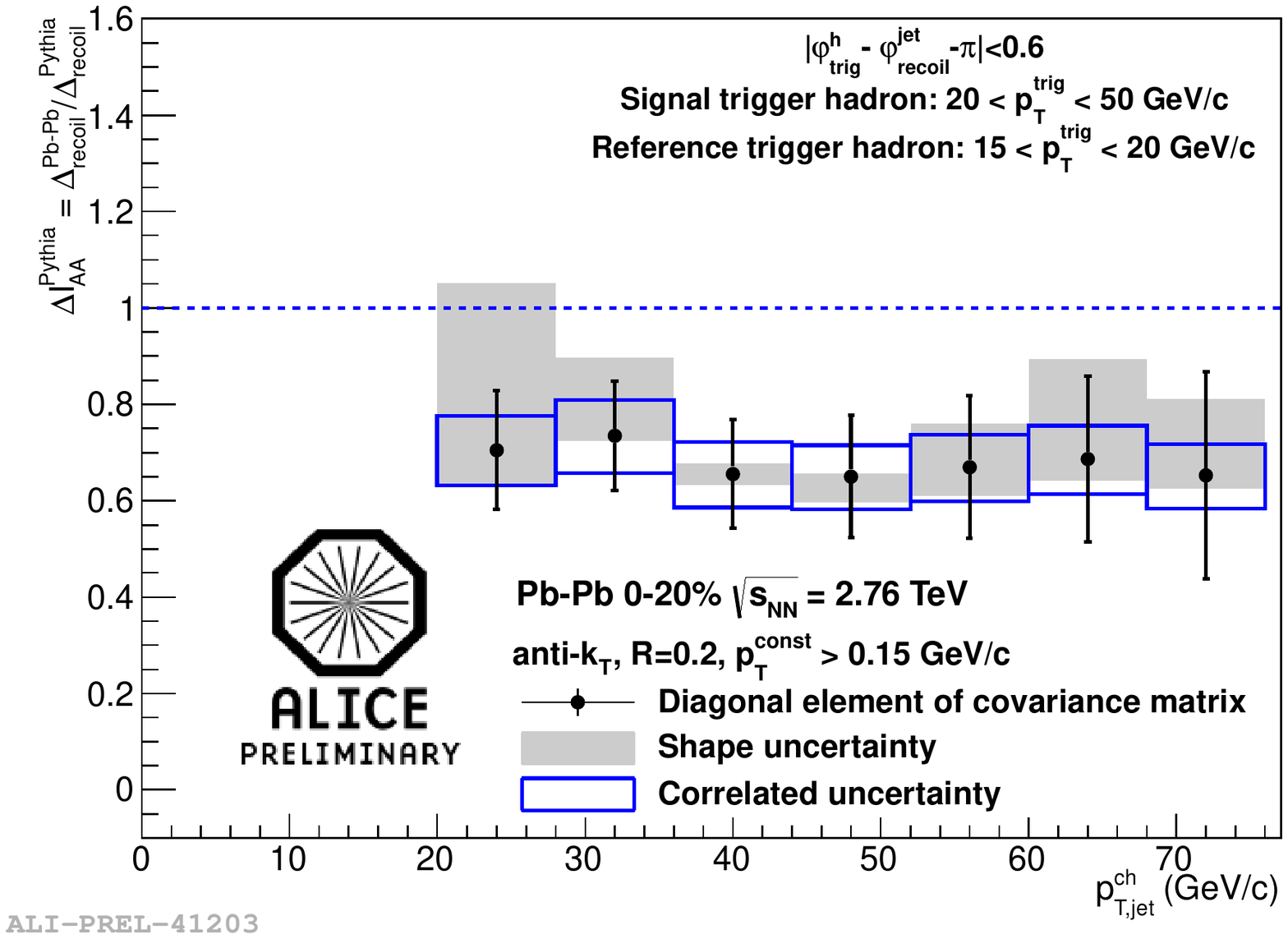}\end{center} \vspace{-4cm}\begin{center}
\includegraphics[width=0.48\textwidth]{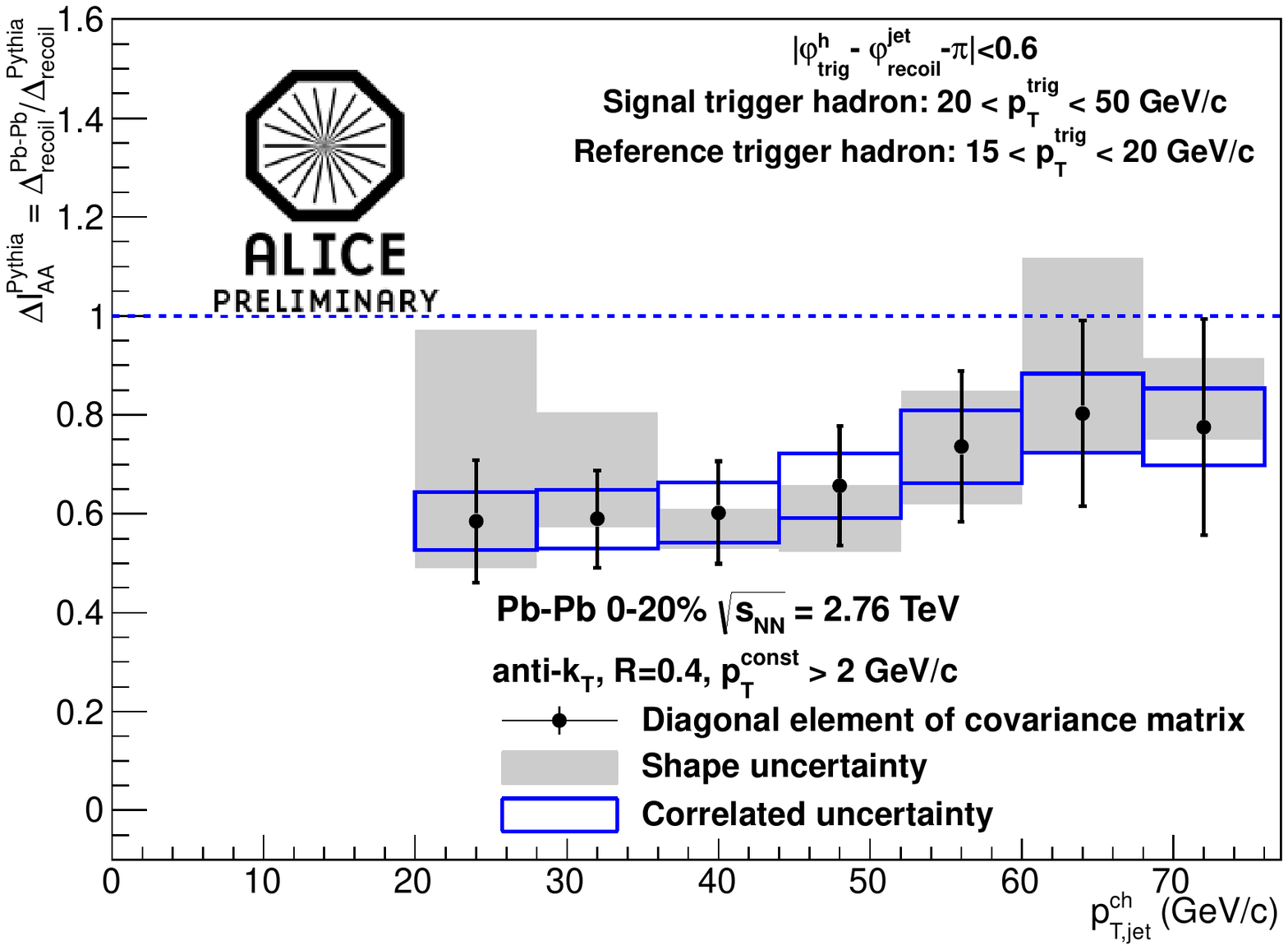}
\end{center}
\caption{$\Delta_{IAA}^{PYTHIA}$ distributions  for different radius and minimum constituent cut selections.}
\label{fig:deltaiaa}
\end{figure}

To explore the energy redistribution within the recoil jets, we
consider the ratio for the measured $\Delta_{recoil}$ distribution
over the same observable calculated with PYTHIA,
 $\Delta_{\rm IAA}^{\rm PYTHIA}$. This ratio is presented in
Fig.\ref{fig:deltaiaa} for $R=0.4$ and $p_{\rm T}^{\rm const} >0.15$ GeV, for
$R=0.2$ and $p_{\rm T}^{\rm const}> 0.15$ GeV and for $R=0.4$ and $p_{\rm T}^{\rm const}> 2$
GeV. Comparison of these distributions does not indicate a large
energy redistribution, relative to PYTHIA, either transverse to the jet axis,
or towards lower \pT\ of constituents, though more precise statements will be possible with reduced systematic uncertanties and higher statistics data.

\vspace{-4cm}
%\section*{References}

\end{document}